\documentclass[prl,twocolumn,showpacs,amsmath,amssymb]{revtex4}

\usepackage{graphicx}
\usepackage{epsfig}
\usepackage{amsmath}
\usepackage{graphics}

\begin{document}

\title{Anomalous Scattering of Low-lying Excitations in a Spin-1 Bose--Einstein Condensate}

\author{Shohei Watabe,$^{1}$} 
\altaffiliation{Present Address: 
Department of Physics, Keio University, Yokohama 223-8522, Japan}
\altaffiliation{CREST(JST), Saitama 332-0012, Japan}
\author{Yusuke Kato$^{2}$ }
\affiliation{
$^{1}$
Department of Physics, The University of Tokyo, Tokyo 113-0033, Japan}
\affiliation{
$^{2}$
Department of Basic Science, The University of Tokyo 153-8902, Japan}

\begin{abstract}
We present the simplest theory of perfect tunneling of an excitation in a Bose--Einstein condensate (BEC) through an impurity potential with an arbitrary shape in the low-momentum limit. 
That is for the transverse spin wave in the ferromagnetic phase of a spin-1 BEC. 
This mode obeys a Schr\"odinger-type equation; yet, 
effects of the potential on its transmission coefficient $T$ and on its scattering cross section $\sigma$ vanish in that limit. 
The order parameter determines $T$, and the momentum $p$-dependence of $\sigma$ exhibits a Rayleigh scattering type ($\sigma \propto p^{4}$). 
These properties are common between two types of Nambu--Goldstone modes: this spin wave and the Bogoliubov mode. 
\end{abstract}

\maketitle

{\it Introduction.---} 
Tunneling through a barrier without reflection is a curious phenomenon. 
Such anomalous behavior is observed in superfluids and superconductors. 
In the early 2000s, 
perfect transmission in the low-momentum limit against a potential barrier 
was predicted in~\cite{Kovrizhin2001,Kagan2003} for the low-lying excitation (the so-called Bogoliubov mode~\cite{Bogoliubov1947}) in a Bose--Einstein condensate (BEC).
This perfect transmission was termed anomalous tunneling~\cite{Kagan2003}. Subsequently, many interesting tunneling properties have been discovered in scalar BECs~\cite{Danshita2006,Kato2008,Fujita2007,Watabe2008,Ohashi2008,Zapata2009,Takahashi2009,Tsuchiya2009}.

In quantum electrodynamics, it is known that a relativistic particle tunnels through a high and wide barrier, which is referred as the Klein paradox~\cite{Kelin1929}; it is also a recent subject of graphene~\cite{Katsnelson2006}. 
This counterintuitive phenomenon is described by the Dirac equation for two-component wave functions (WFs) and its particle has the linear dispersion. 
Since the Bogoliubov mode has the linear dispersion in the low-momentum limit and obeys the Bogoliubov equation for the two-component WFs, 
one might expect that this paradox is linked to anomalous tunneling of the Bogoliubov mode via some ways like the charge-conjugation symmetry for the Dirac equation.

Two years ago, however, 
anomalous tunneling through a generic barrier 
was derived using the coincidence between the WF of the Bogoliubov mode in the low-energy limit and that of the condensate~\cite{Kato2008}, which is regarded as a property of NG modes. 
With this explanation, we find ourselves confronted with the following questions which we take up below: 
Is anomalous tunneling inherent to Nambu--Goldstone (NG) modes in broken-symmetry states? 
Does a NG mode with a parabolic dispersion relation described by a Schr\"odinger-type equation (corresponding {\it e.g.} to the ferromagnetic magnon) exhibit anomalous scattering properties?

The ferromagnetic state of the spin-1 spinor BEC~\cite{Ohmi1998,Ho1998}
(a BEC composed of particles with spin-1 internal degrees of freedom)
 is suitable for investigating these unresolved questions since both gauge and spin rotational symmetries are spontaneously broken and accordingly the system hosts two types of NG modes: 
the Bogoliubov mode (type I) and the spin wave (type II)~\cite{NoteType}. 
By comparing the scattering properties of these two modes, we can gain a deeper understanding of the scattering properties inherent to NG modes in broken-symmetry states. 
Further, the study of the spin wave in the spinor BEC is helpful for understanding the scattering properties of spin waves in magnetic systems; 
tunneling of spin waves in a magnetic film through an inhomogeneous magnetic field was recently demonstrated~\cite{Demokritov2004, Hansen2007}.

With this background, 
we investigate
the scattering properties of low-lying excitations in the ferromagnetic phase of a spin-1 BEC from an obstacle within mean-field theory. 
We examine their tunneling properties and their cross section $\sigma$ of scattering from a generic potential with spherical symmetry. 
Our main findings on the spin waves are that anomalous tunneling occurs at the low-energy limit, and that the Rayleigh scattering type $k$-dependence $\sigma\propto k^4$ holds at small wavenumbers $k$.
Comparison with the Bogoliubov mode and inspection of the derivation suggest that both findings hold for other NG modes regardless of the spectrum.

{\it Formulation.---} 
We start with a Hamiltonian for spin-1 bosons with the atomic mass $m$, given by~\cite{Ho1998} 
\begin{align}
\hat{\mathcal H} = & \int d{\bf r} \left [ \frac{\hbar^{2}}{2m} \nabla \hat{\Psi}_{j}^{\dag} \nabla \hat{\Psi}_{j} 
+ V_{\rm ext} \hat{\Psi}_{j}^{\dag} \hat{\Psi}_{j} 
\right . 
\nonumber 
\\
& 
+ \frac{c_{0}}{2} (\hat{\Psi}_{j}^{\dag} \hat{\Psi}_{j})^{2} 
\left. 
+ \frac{c_{1}}{2} (\hat{\Psi}_{j}^{\dag} {\bf S}_{jj'} \hat{\Psi}_{j'})^{2}
\right ], 
\label{eq: Hamiltonian}
\end{align}
where $\hat{\Psi}_{j}$ is the field annihilation operator for a spin-1 boson in hyperfine state $j$ $(j = \pm1, 0)$, 
$S_{\alpha}$ $(\alpha = x,y,z)$ are spin matrices, and repeated indices are summed. 
When we denote by $a_{F}$ the $s$-wave scattering length for the total spin $F$ channel, 
two coupling constants are given by $c_{0} \equiv  4\pi\hbar^{2}(a_{0}+2a_{2})/(3m)$ and $c_{1} \equiv 4\pi\hbar^{2}(a_{2}-a_{0})/(3m)$. 
The external potential $V_{\rm ext}$ is coupled only to the local density, 
so that 
$\hat{\mathcal H}$ 
is $U(1)\times SO(3)$ invariant~\cite{Ho1998}. 

The mean-field theory of the spin-1 BEC was established in~\cite{Ohmi1998,Ho1998}. 
The ground state of the ferromagnetic phase is realized for $c_{1} < 0$, 
and its condensate WFs (order parameters) $\langle \hat{\Psi}_{j} \rangle \equiv \Phi_{j}e^{-i\mu t/\hbar}$ are given by $(\Phi_{+1}, \Phi_{0}, \Phi_{-1}) = (\sqrt{n} \phi ({\bf r}), 0, 0)$~\cite{Ho1998}, 
where $\mu$ is the chemical potential and $n$ is the density of the condensate in the uniform regime where $\phi ({\bf r})$ becomes unity. 
We note that $\phi$ obeys an equation which is identical in form with the GP equation of a scalar BEC: 
$\hat{\mathcal H}_{\rm GP}  \phi=0$, 
with $\hat{\mathcal H}_{\rm GP} \equiv - (\hbar^{2}\nabla^{2})/(2m) + V_{\rm ext} + c_{+}n (\phi^{2} - 1)$ 
and $c_{+} \equiv c_{0} + c_{1}$, where we used $\mu = c_{+}n$~\cite{Ohmi1998}. 
In the ground state, we can take $\phi$ as real.
This system includes three types of excitations~\cite{Ohmi1998,Ho1998}. 
The excitation of $j=+1$ is mathematically identical to the Bogoliubov mode of the scalar BEC; 
its tunneling properties can be drawn from results that have been established 
in previous studies~\cite{Kovrizhin2001,Kagan2003,Danshita2006,Kato2008,Fujita2007,Watabe2008,Ohashi2008,Zapata2009,Takahashi2009,Tsuchiya2009}. 
The spectrum of the mode $j=-1$ is massive: $E = \hbar^{2}k^{2}/(2m)  + 2 |c_{1}| n$.  
Since our emphasis is on the nature of NG modes, we will not discuss further this massive excitation in this Letter. 
Its properties have been partially reported in~\cite{WatabeKato2009} and will be reported elsewhere. 
The mode of $j=0$ leads to the transverse spin wave with the spectrum $E = \hbar^{2}k^{2}/(2m)$, 
which we focus on in this study. 
When $\tilde{\phi}_{0}$ denotes the WF of the excitation for $j = 0$, 
its equation of motion is given by~\cite{Ho1998} 
\begin{align}
i\hbar \frac{\partial \tilde{\phi}_{0}}{\partial t}
= & 
\hat{\mathcal H}_{\rm GP} 
\tilde{\phi}_{0}, 
\quad\mbox{with}\quad\tilde{\phi}_{0} \equiv \phi_{0}\exp{(-iE t/\hbar)}. 
\label{Eq8}
\end{align} 
We introduce dimensionless units: $\bar{E}\equiv E/(c_{+}n)$, $\bar{V}_{\rm ext}\equiv V_{\rm ext}/(c_{+}n)$, 
$\bar{k} \equiv k\xi_{\rm f}$, and $\bar{x} \equiv x/{\xi_{\rm f}}$ where $\xi_{\rm f} \equiv \hbar\sqrt{mc_{+}n}$. 
We hereafter omit the bar for simplicity. 
 
{\it Transmission coefficient.---} 
We first study reflection and refraction of the transverse spin wave. 
We consider a potential $V_{\rm ext} ({\bf r}) = V_{\rm ext}(x)$ with only an $x$ dependence, 
where $V_{\rm ext}(x)$ acts as a potential wall near $x\sim 0$. 
We further impose the following condition: $V_{\rm ext}(x) \rightarrow V_{\rm R}$ for $x \gg 1$, and $V_{\rm ext}(x) \rightarrow V_{\rm L}$ for $x \ll - 1$.
We assume that the incident wave with momentum $k$ propagates at an angle $\theta_{\rm L}$ with respect to the $x$ axis
~\cite{IncidentAngle}. 
For $|x| \rightarrow \infty$, we find $\phi \rightarrow \sqrt{1-V_{\nu}} (\equiv A_{\nu})$ through the equation of $\phi$, 
where $A_{\nu}$ is the amplitude of the order parameter $|\Phi_{+1}| (= |\langle \hat{\Psi}_{+1} \rangle|)$, 
and the subscript $\nu$ denotes L (or R) for the asymptotic regime of $x <0 \, ({\rm or} >0)$. 
In dimensionless form, 
the healing length is modified such that $\xi_{\rm L (or R)} \equiv 1/A_{\rm L (or R)}$ for the regime $x < 0\, ({\rm or} >0)$, 
since the condensate density is not unity at $|x|\rightarrow \infty$. 
Even if $V_{\rm L} \neq V_{\rm R}$, 
there is a common energy spectrum $E=\hbar^{2}k^{2}/(2m)$ for $|x| \gg \xi_{\rm L, R}$ from (\ref{Eq8}). 
From this result and the continuity of the WF, it follows that the spin wave is not refractive. 
This is to be contrasted with the behavior of the Bogoliubov mode, which was shown to be refractive in~\cite{Watabe2008}, obeying Snell's law.

We take the $xy$ plane as the incident plane without loss of generality, 
and set $\phi_0({\bf r})=X(x)\exp(iky\sin\theta_{\rm L})$. 
Let $k_{x} \equiv k\cos{\theta_{\rm L}}$ be the $x$-component of the momentum, and 
let us define $\hat{H}_{\rm GP} \equiv -\frac{1}{2} \frac{d^{2}}{dx^{2}}+ V_{\rm ext} + \phi^{2}-1$. 
For elucidating tunneling properties, we solve
\begin{equation}
  \hat{H}_{\rm GP}X(x) = 
\frac{k_x^2}{2} X(x), 
\label{eq: eqforX}
\end{equation} 
with the following boundary condition: 
\begin{equation}
X(x)
= \left \{ 
\begin{array}{ll}
\exp{(i k_{x} x)} + r \exp{(-i k_{x} x)} \, &  {\rm for} \,\, x \ll -\xi_{\rm L}, 
\\
t \exp{(i k_{x} x)} \, & {\rm for} \,\, x \gg \xi_{\rm R}.
\end{array}
\right .
\label{Eq14}
\end{equation}
We denoted by $r=r(k_x)$ and $t=t(k_x)$ respectively the amplitude reflection and transmission coefficients.
The transmission and reflection coefficients are given by $T(k) = |t(k)|^{2}$ and $R(k) = |r(k)|^{2}$. 

Below, we prove that 
these coefficients are given by
\begin{align}
\lim\limits_{k\rightarrow 0}T(k) = \frac{4M_{\rm L}M_{\rm R}}{(M_{\rm L} + M_{\rm R})^{2}}, 
\lim\limits_{k\rightarrow 0}R(k) = \frac{(M_{\rm L} - M_{\rm R})^{2}}{(M_{\rm L} + M_{\rm R})^{2}}. 
\label{Result1}
\end{align} 
Here, $M_{\nu}$ for $\nu = {\rm L}$ and ${\rm R}$ denotes the modulus of the magnetization $|{\bf M}|$ in each asymptotic regime; it is expressed as $\mu_{\rm B}nA_{\nu}^{2}$ when we recover the conventional units. 
This result satisfies $T+R=1$, 
as well as the usual reciprocity relationship for tunneling.  When $V_{\rm L}=V_{\rm R}$, 
the magnetizations on the left and right sides are equal (i.e., $M_{\rm L}=M_{\rm R}$) and perfect transmission at low energy 
$T(k\rightarrow 0)=1$ 
follows regardless of the forms and heights of the barriers about $x=0$. 
This is the simplest example of anomalous tunneling; 
earlier anomalous tunneling is derived through a two-component Schr\"odinger equation for the Bogoliubov mode~\cite{Kovrizhin2001,Kagan2003}, 
while anomalous tunneling of this spin wave is derived through {\it a single-component} Schr\"odinger equation. 
We find that this tunneling phenomenon is nontrivial by recalling that 
a particle obeying the {\it single-component} Schr\"odinger equation 
generally undergoes total reflection against a potential barrier at $k\rightarrow 0$. 

Let us prove (\ref{Result1}). 
We first seek the solution to (\ref{eq: eqforX}) in the form of $ X(x) = \sum_{n=0}^{\infty} k_{x}^{2n} X^{(n)}(x)$. 
$X^{(n)}(x)$ satisfy
\begin{align}
\hat{H}_{\rm GP}X^{(0)}(x)=0,\quad
\hat{H}_{\rm GP}X^{(n)}(x) = 
\frac{1}{2} X^{(n-1)}(x) 
\label{eq: resursion}
\end{align} 
for $n\ge 1$. As a set of linearly independent solutions to the first equation of (\ref{eq: resursion}), we take 
$
X_{\rm I}(x) \equiv \phi (x)$ and 
$
X_{\rm II}(x) \equiv 
- \gamma \phi(x) 
+ \phi (x) \int_{0}^{x}dx' {\phi^{-2}(x')}$. 
Here, $\gamma \equiv (\gamma_{\rm R} + \gamma_{\rm L})/2$, 
$
\gamma_{\rm L} \equiv 
\int_{0}^{-\infty} dx \left ( {\phi^{-2} (x)} - {A_{\rm L}^{-2}} \right ), 
$
and 
$
\gamma_{\rm R} \equiv 
\int_{0}^{\infty} dx \left ( \phi^{-2} (x) - A_{\rm R}^{-2} \right) 
$. 
In the asymptotic regime $x \ll -\xi_{\rm L}$ ($x \gg \xi_{\rm R}$), we have 
$X_{\rm I}(x) \sim A_{\rm L (R)}$, and 
$X_{\rm II}(x) \sim  x/{A_{\rm L (R)}}+ {\rm sgn}(x) \gamma_- A_{\rm L (R)}$, 
with 
$\gamma_-=(\gamma_{\rm R}-\gamma_{\rm L})/2$. 
The solution to (\ref{eq: eqforX})
can be written as 
\begin{align} 
X(x;k_x)= C_{\rm I} (k_x) X_{\rm I}(x;k_x^2)+C_{\rm II} (k_x) X_{\rm II}(x;k_x^2) 
\label{Eq11}
\end{align}
with $X_{\alpha}(x;k_x^2)\equiv \sum_{n=0}^{\infty} k_{x}^{2n} X_{\alpha}^{(n)}(x)$ for $\alpha={\rm I}, {\rm II}$ and 
\begin{align}
X_{\alpha}^{(0)}(x)= & X_\alpha(x),\\
X_{\alpha}^{(n)}(x)= & -\phi(x)\int_{0}^{x}\frac{dx'}{\phi^{2}(x')}\int_{0}^{x'}dx'' \phi(x'')  X_{\alpha}^{(n-1)}(x''), 
\end{align}
for $n\ge 1$, where 
$C_{\alpha}(k_x)$ are coefficients. 
Equating (\ref{Eq14}) with (\ref{Eq11}) in the asymptotic regions $\xi_{\rm L,R}\ll |x|\ll |k_x|^{-1}$, 
we obtain 
$\lim_{k\rightarrow 0}t(k) = 2A_{\rm L}A_{\rm R}/(A_{\rm L}^{2} + A_{\rm R}^{2})$
and $\lim_{k\rightarrow 0}r(k) = (A_{\rm L}^{2} - A_{\rm R}^{2})/(A_{\rm L}^{2} + A_{\rm R}^{2})$. 
We can then arrive at (\ref{Result1}) with $T(k) = |t(k)|^{2}$ and $R(k) = |r(k)|^{2}$.

In the derivation of (\ref{Result1}), the crucial facts were (i) that the WF of the excitation in the low-momentum limit is proportional to the condensate WF 
and (ii) that the healing length for the condensate is finite so that the condensate WF recovers the value in the spatially uniform system from the depleted value around the potential over a finite transient region.

The Hamiltonian (\ref{eq: Hamiltonian}) is invariant under gauge transformation and spin space rotation, since $V_{\rm ext}$ is coupled only to the local density (i.e., symmetry-preserving potential). 
When continuous symmetry is spontaneously broken, the ground state has infinite degeneracy. 
Even in the spatially inhomogeneous system, the WF of the corresponding excitation at $E\rightarrow 0$ describes the NG mode. 
The fact that the spin wave is a NG mode is therefore linked to (i). 
The property (i) is unusual, since 
the WF obeying the Schr\"odinger equation equals the order parameter of an interacting Bose gas.

As for (ii), the finite healing length is characteristic of the interacting Bose gas; this interaction yields finite compressibility. 
When the compressibility is finite and (i) holds, the WF of the NG mode at the low-momentum limit as well as the condensate WF persist with finite and nonzero amplitudes near the potential barrier. 
This property of the WF leads to the nonzero transmission of the NG mode in the low-momentum limit. 

{\it Scattering cross section.---} 
We investigate the scattering cross section of the transverse spin wave 
from the spherical potential $V_{\rm ext}({\bf r}) = V (r)$, 
which satisfies $\lim_{r\rightarrow 0}r^2 V (r) =0$ and 
$\lim_{r\rightarrow \infty}V(r)=0$. 
In the ground state, the condensate WF is also spherically symmetric: $\phi ({\bf r})=\phi (r)$.
The GP equation in dimensionless form is reduced to $\hat{\mathcal H}_{r} \phi (r) = 0$ with 
$\hat{\mathcal H}_{r} \equiv - [(d^{2}/dr^{2}) +(2/r)(d/dr)]/2 + \tilde{V}(r)$ and $\tilde{V}(r)\equiv V(r) + \phi^{2}(r) -1$. 
$\phi_{0} ({\bf r})$ can be expanded with respect to the partial waves
$
\phi_0({\bf r})=\sum_{l=0}^\infty P_l(\cos\theta)R_l(r;k)$ with
the Legendre polynomial $P_l(x)$ and the angle $\theta$ between ${\bf r}$ and the incident momentum.
The radial WF $R_l(r;k)$ satisfies 
\begin{align}
\left [\hat{\mathcal H}_{r} + \frac{l(l+1)}{2 r^{2}} \right ]R_l(r;k)=\frac{k^{2}}{2} R_l(r;k), 
\end{align}
with $E=k^2/2$. 
It behaves as $R_l(r;k)\propto r^l$ for $r\sim 0$ and 
$R_l(r;k)\propto j_l(kr)-\tan\delta_l(k)n_l(kr)$ for $r\rightarrow \infty$. 
Here, $j_{l}(kr)$ and $n_{l}(kr)$ respectively denote the spherical Bessel and Neumann functions, 
and $\delta_l(k)$ denotes the phase shift of the partial wave. 

If $\tilde{V}(r)$ were a generic potential, the phase shifts would behave as $\delta_l(k)\propto k^{2l+1}$ for small $k$ and 
the cross section $\sigma(k)=4 \pi k^{-2}\sum_{l=0}^\infty (2l+1) \sin^2\delta_l(k)$ would be given by
\begin{equation}
\sigma(k)\sim k^{-2}\sin^2\delta_0(k)\sim \mbox{const.}(>0) 
\label{normalcrosssection}
\end{equation}
for small $k$. 
However, in the present case, even when $V(r)$ is generic, 
$\tilde{V}(r)$ reflects the specific properties of the condensate WF. 
In the following few paragraphs, we show that this aspect has a significant influence on the $s$-wave scattering behavior, leading, at small $k$, to 
\begin{equation}
\tan\delta_0(k)\propto k^3,\quad\sigma(k)\propto k^4. 
\label{eq: tansigma}
\end{equation}
The derivation of (\ref{eq: tansigma}) relies on the following two relations which correspond to (i) and (ii) noted previously: 
\begin{eqnarray}
\lim_{k\rightarrow 0}R_0(r;k)\propto \phi(r), 
\label{eq: condition1}
\\
\lim_{r\rightarrow \infty}r^m(\phi(r)-1)=0,
\label{eq: condition2}
\end{eqnarray}
for $m\ge 0$. 
In what follows, we fix the overall normalization factor of $R_0(r;k)$ such that 
\begin{equation}
R_0(r=0;k)=\phi(r=0). 
\label{eq: bcatr0}
\end{equation}

Let us expand $R_0(r;k)$ as $r^{-1}\sum_{n=0}^\infty k^{2n}u^{(n)}(r)$ with respect to $k^2$. It suffices to retain the first two terms $u^{(0)}(r)$ and $u^{(1)}(r)$ to prove (\ref{eq: tansigma}). 
$u^{(0)}(r)$ and $u^{(1)}(r)$ respectively satisfy 
\begin{align}
\hat{\mathcal H}'_{r} u^{(0)}(r)=0,\quad
\hat{\mathcal H}'_{r} u^{(1)}(r)=u^{(0)}(r)/2, 
\label{eq: u0u1}
\end{align}
where $\hat{\mathcal H}'_{r} \equiv - \frac12\frac{d^2}{dr^2} + \tilde{V}(r)$. 
A set of linearly independent solutions to the first equation of (\ref{eq: u0u1}) is given by 
$u_{\rm I}(r)\equiv r \phi(r)$ and $u_{\rm II}(r)\equiv u_{\rm I} (r)\int^r_\infty (u_{\rm I} (r'))^{-2}dr'$. 
$u_{\rm I}(r)$ is obtained by noting (\ref{eq: condition1}). 
The condition (\ref{eq: bcatr0}) yields $u^{(0)}(r)=u_{\rm I} (r)$, 
since the second solution behaves as $u_{\rm II} (r)\sim r^0$ near $r=0$. 
A special solution to the second equation of (\ref{eq: u0u1}) is, on the other hand, given by
\begin{equation}
u^{(1)}(r)=u_{\rm I} (r)\int^r _0 u_{\rm I} (r')u_{\rm II} (r')dr'-u_{\rm II} (r)\int^r _0 (u_{\rm I} (r'))^2 dr'.
\label{eq: u1}
\end{equation}
With use of (\ref{eq: u1}) and $u^{(0)}(r)=u_{\rm I}(r)$, 
we examine the asymptotic behavior of 
\begin{equation}
R_0(r;k)=u^{(0)}(r)/r + k^{2}u^{(1)} (r)/r + {\cal O}(k^4),  
\label{eq: R0}
\end{equation}
at $r\rightarrow \infty$. 
(\ref{eq: condition2}) yields $u^{(0)}(r)/r=u_{\rm I} (r)/r\sim 1$ and $u^{(1)}(r)/r \simeq - r^{2}/{6} + B + C/r$ at $r\rightarrow \infty$. 
Here, $B$ and $C$ are given by
\begin{equation}
B \equiv \int_0^\infty [u_{\rm I} (r')u_{\rm II} (r')+r']dr',\quad
C \equiv \int_0^\infty [(u_{\rm I} (r'))^2-r'^2]dr'.
\end{equation} 
The asymptotic form of $R_0(r;k)$ then reads 
\begin{equation}
R_0(r;k)=
(1+B k^{2}) \left ( 1-\frac{k^{2}r^{2}}{6} \right) - C k^{3} \left ( \frac{-1}{kr} \right ) + \cdots. 
\end{equation} 
Since $j_{0}(kr)$ and $n_{0}(kr)$ are respectively given by $j_{0}(kr)\simeq 1- k^{2}r^{2}/6$ 
and $n_{0}(kr) \simeq -1/(kr)$ for $kr \ll 1$, 
the phase shift of the $s$-wave ($l=0$) can be written as 
$\tan \delta_{0} \simeq Ck^{3}/(1+B k^{2})\propto k^{3}$. 
(\ref{eq: tansigma}) has been proved. 

There are known examples in which $\sigma(k)\propto k^4$ holds for the wave with $E\propto k$; 
these examples include sound, light (Rayleigh scattering), and the Bogoliubov mode~\cite{Kato2008,Fujita2007} of a BEC. 
Our result $\sigma(k)\propto k^4$ for low-energy scattering of the spin mode with $E \propto k^{2}$ is rather unfamiliar compared to (\ref{normalcrosssection}) 
and is regarded as an anomalous power law. 

{\it Discussion.---} 
The transmission coefficient $T$ of the Bogoliubov mode in the spin-1 BEC is the same as that in the scalar BEC; 
we here recall this coefficient of the Bogoliubov mode in the low-energy limit~\cite{Watabe2008, Ref19} given by 
\begin{align}
\lim\limits_{k\rightarrow 0}T = 
\frac{4A_{\rm L}A_{\rm R}\cos{\theta_{\rm L}}\cos{\theta_{\rm R}}
}{
(A_{\rm L}\cos{\theta_{\rm L}}+ A_{\rm R}\cos{\theta_{\rm R}})^{2}
}, 
\label{eq: Tbogo}
\end{align} 
where $\theta_{\rm R}$ is the refracted angle and others are defined previously. 
Comparing (\ref{Result1}) and (\ref{eq: Tbogo}) with the transmission coefficient of classical waves, we see that the modulus of magnetization $M_\nu$ and that of the condensate WF $A_\nu$ both play the role of ^^ ^^ impedance". 
We can then extract the following common feature of both low-lying modes: 
(a) 
In a system whose continuous symmetry is spontaneously broken, 
impedance of the corresponding NG mode in the transmission coefficient is given by the amplitude of its order parameter
({\it e.g.}, $|\langle \hat\Psi \rangle|$ 
or 
$|{\bf M}|$). 
We omitted the subscript $j=+1$ of $\hat{\Psi}_{j}$. 
As for the low-energy scattering from a spherical potential, we previously noted another unexpected feature common to the two low-lying excitations: 
(b) 
the $s$-wave scattering cross section is proportional to the fourth power of the momentum. 
Table~\ref{table1} summarizes the scattering properties of the two NG modes. 
The derivations of (a) and (b) both rely on (i) and (ii), and are straightforward and general without use of any other property specific to the system. 
We thus anticipate that (a) and (b) both hold as general properties of NG modes in the presence of symmetry-preserving potentials~\cite{Polar}. 

\begin{table}[htbp]
\begin{center}
\caption{\label{table1} 
Scattering properties of NG modes in the ferromagnetic phase of the spin-1 spinor BEC}
\begin{tabular}{cccccc}
\hline\hline
NG mode     & \shortstack{type of \\ NG mode} & refraction & impedance & \shortstack{scattering \\ cross section}
\\
\hline
\shortstack{Bogoliubov}     & \shortstack{I} & \shortstack{Snell's law}  & \shortstack{$|\langle \hat\Psi \rangle|$} & \shortstack{$\sigma_{0} \propto k^{4}$}
\\
\shortstack{transverse spin}     & \shortstack{II} & \shortstack{no}  & \shortstack{$| {\bf M} |$} & \shortstack{$\sigma_{0} \propto k^{4}$}
\\
\hline\hline
\end{tabular}
\end{center}
\end{table}


It was demonstrated that spin waves in magnetic systems transmit a barrier induced 
by the inhomogeneous magnetic field~\cite{Demokritov2004}. 
This high tunneling probability is due to the nonlocal magnetic dipole interaction of spins. 
We note that this is different from our situations; 
we propose a completely different mechanism of high tunneling probability through the {\it nonmagnetic} barrier. 
It stems from the degeneracy of the ground state induced by the spontaneous breaking of the continuous symmetry. 
In magnetic systems whose nonlocal magnetic dipole interaction is negligible, 
the following two $s$-wave scattering cross sections will be observed: 
the first is the Rayleigh type (\ref{eq: tansigma})
for a doped nonmagnetic impurity (e.g., a closed shell atom) or for a defect like a mechanical depression of the magnetization. 
The second is a usual type (\ref{normalcrosssection}) for a doped magnetic impurity (e.g., an atom classified as transition metals or rare earth elements). 
Propagation of the spin waves can be manipulated with impurities and defects adequately disposed. 
It was recently reported that electric signals in a magnetic insulator are transported by converting the electric current to a spin wave~\cite{Kajiwara2010}. 
If (\ref{eq: tansigma}) holds and there are no magnetic impurities, 
the scattering of spin waves from nonmagnetic impurities and inhomogeneities is suppressed for lower momentum, 
and electric signals could be transported over longer distances at lower energy. 

Note that, in ultracold gases, the diffraction and the scattering problem of excitations from a spherical potential are now possible to investigate if one uses a single ion trapped inside a BEC~\cite{Zipkes2010}. 

{\it Conclusion.---} 
We constructed the simplest model of an anomalous tunneling phenomenon 
for an excitation through a generic localized potential, 
using a transverse spin wave in the ferromagnetic phase of a spin-1 BEC. 
Interestingly, studies of anomalous tunneling will connect to many systems 
whose continuous symmetry is spontaneously broken. 
Our present study will be a minimal model for these studies. 

%

{\it Acknowledgement.---} 
We thank A. Tanaka and R. Yamashita for their helpful comments on this manuscript. 
This work is supported by KAKENHI (21540352) from JSPS and (20029007) from MEXT in Japan. 
S. W. was supported by a Grant-in-Aid for JSPS Fellows (217751).


\begin{thebibliography}{99}
\bibitem{Polar}
We verified that all excitations in the polar phase of a spin-1 BEC exhibit anomalous tunneling, and (i) and (ii) both hold.
\end{thebibliography}
\end{document}